\documentclass[aps,pra,epsfigure,twocolumn,longbibliography]{revtex4-1}
\usepackage{dcolumn}    
\usepackage{bm} 
\usepackage{graphicx}
\usepackage{amsmath}    
\usepackage{latexsym}
\usepackage{amsfonts}   
\usepackage{amssymb}
\usepackage{array}      
\usepackage{epsfig}
\usepackage{txfonts}
\usepackage{color}
\usepackage[colorlinks=true,linkcolor=blue,urlcolor=blue,citecolor=blue,pdfusetitle]{hyperref}
\usepackage{hyperref}

\newcommand{\ket}[1]{\left\vert#1\right\rangle}
\newcommand{\bra}[1]{\left\langle#1\right\vert}
\newcommand{\blah}{blah\\blah\\blah\\blah\\blah.}

\definecolor{DarkGreen}{rgb}{0.00,0.39,0.00}

\catcode`\|=\active \def|{
\fontencoding{T1}\selectfont\symbol{124}\fontencoding{\encodingdefault}}
\newcommand{\mathd}{\mathrm{d}}
\newcommand{\tmop}[1]{\ensuremath{\operatorname{#1}}}

\begin{document}
 \title{Entropy production and correlations in a controlled non-Markovian setting}
       
\author{Maria Popovic,$^1$, Bassano Vacchini$^{1,2}$, and Steve Campbell$^{2}$}
\affiliation{$^1$Dipartimento di Fisica ``Aldo Pontremoli", Universit{\`a} degli Studi di Milano, via Celoria 16, 20133 Milan, Italy\\
$^2$Istituto Nazionale di Fisica Nucleare, Sezione di Milano, via Celoria 16, 20133 Milan, Italy}

\begin{abstract}
We study the relationship between (non-)Markovian evolutions, established correlations, and the entropy production rate. We consider a system qubit in contact with a thermal bath and in addition the system is strongly coupled to an ancillary qubit. We examine the steady state properties finding that the coupling leads to effective temperatures emerging in the composite system, and show that this is related to the creation of correlations between the qubits. By establishing the conditions under which the system reaches thermal equilibrium with the bath despite undergoing a non-Markovian evolution, we examine the entropy production rate, showing that its transient negativity is a sufficient sign of non-Markovianity.
\end{abstract}
\date{\today}
\maketitle

\section{Introduction}
The inevitable interaction of a system with its surroundings necessitates we find suitable means for modelling the dynamics of open quantum systems~\cite{Breuer2002}. Under certain situations, in particular when the coupling between the system and its environment is sufficiently weak such that the environment is unaffected by this interaction, the system evolves in a Markovian (memoryless) manner. For these settings, a well-known and widely used approach is to model the time evolution using a dynamical semi-group. Conversely, allowing for memory effects the scenario drastically changes~\cite{CampbellPRA2012}, and different approaches to the very definition of a non-Markovian dynamics have been recently introduced~\cite{Vacchini2011a,Vacchini2012a,AddisPRA}, as discussed in the recent reviews~\cite{Rivas2014a,BreuerRMP}.

An equally important issue is developing a clear thermodynamic framework for quantum systems~\cite{GooldJPA,Alicki2018a}. Indeed, thermodynamic quantities, such as work and heat, must be carefully re-examined when the working materials are inherently quantum. In the thermodynamic characterization of a given process, the (irreversible) entropy production and the associated entropy production rates are crucial~\cite{Spohn, Alicki1979, EspositoNJP, DeffnerPRL2011, DeffnerPRE2015}. The entropy production can be naturally defined as the difference between the change in entropy of the reduced system state and the mean exchanged heat with a bath at fixed temperature, $T$, divided by $T$. For the case of a quantum dynamical semigroup with a stationary state in Gibbs form the entropy production is guaranteed to be positive or zero, and can naturally be associated with a statement of the second law. An equivalent expression for the entropy production in a semigroup dynamics can be introduced also in the presence of an invariant state~\cite{Spohn}. However, in general, this definition lacks a clear thermodynamic interpretation since the invariant state is not necessarily a thermal equilibrium state. To date several significant advances have been made in defining and understanding the thermodynamic entropy production for quantum systems~\cite{PaternostroPRL2017, PaternostroArXiv, GooldArXiv, BarbieriArXiv, BrunelliArXiv, SerraPRL2016, DeffnerPRE2017}, however only recently has the explicit consideration of non-Markovian maps, where negative entropy production rates can appear, been explored~\cite{MarcantoniSciRep, PatiPRA, ManiscalcoSciRep}.

It is in this direction that the present work progresses. We consider a two-level quantum system (qubit) immersed in a Markovian bath. The system is in turn strongly coupled to an ancillary two-level system such that the joint dynamics of the two qubits is Markovian, while the reduced dynamics of the system alone is manifestly non-Markovian. Thus, our setting significantly differs from other recent works, for example Refs.~\cite{MarcantoniSciRep, PatiPRA}, as we have a direct access to the state of both the system and ancilla. Indeed, in our model we can consider the ancilla as a special subset of environmental degrees of freedom with which the system interacts and which gives rise to a non-Markovian evolution. Thus, at variance with other studies, one of the main goals of the present work is to assess the role that the establishment of correlations plays in the thermodynamic characterization of the evolution. By first characterizing the steady state properties, we show that the strong coupling can lead to a non-equilibrium steady state exhibiting correlations between the two qubits. These correlations can be related to the emergence of effective temperatures, different from that of the bath, for the system and the ancilla, making a thermodynamic description of the process more complex. Despite this, as the overall system+ancilla evolves under a Markovian map, a meaningful, albeit not necessarily thermodynamically meaningful, entropy production can be studied. By identifying the conditions under which the reduced system qubit reaches thermal equilibrium, we then study its associated entropy production and entropy production rate. While the long-time entropy production is consistent with the Markovian case, the coupling induced non-Markovian dynamics can lead to transiently negative rates. While such negative entropy production rates are due to the non-Markovian dynamics, the two notions are not necessarily commensurate, which we show by examining the trace distance measure of non-Markovianity.

\section{The model}
We consider a bipartite system consisting of two coupled spin-$1/2$ particles (qubits), labelled $S$ and $A$. The free evolution of the qubits are governed by their respective Hamiltonians $\mathcal{H}_{S(A)}\!=\!\omega_{S(A)} \sigma_z$. Throughout we will denote $\ket{0}(\ket{1})$ as their ground (excited) state and assume units such that $\hbar=k_B=1$. We will assume that only qubit $S$ feels the effects of an external thermal environment (bath) and thus the dynamics of the total system can be described by the Markovian master equation (omitting the explicit time dependence)~\cite{CampbellPRA2010}
\begin{equation}
\label{master}
\dot{\varrho}_{SA} = \mathcal{L}(\varrho_{SA}) = -i[\mathcal{H}_I +\mathcal{H}_S + \mathcal{H}_A , \varrho_{SA}] + \mathcal{D}(\varrho_{SA}),
\end{equation}
with $\mathcal{H}_I= \left(J_x \sigma_x \otimes \sigma_x + J_y \sigma_y \otimes \sigma_y +J_z \sigma_z \otimes \sigma_z\right)$ defining the interaction between the two qubits and
\begin{equation}
\begin{aligned}
\mathcal{D}(\varrho_{SA}) & = \gamma \left( (\sigma_- \otimes \openone) \varrho_{SA}  (\sigma_+ \otimes \openone) - \frac{1}{2} \Big\{ \varrho_{SA} , (\sigma_+  \otimes \openone) (\sigma_-  \otimes \openone) \Big\} \right) \\ & + \Gamma \left( (\sigma_+ \otimes \openone) \varrho_{SA} (\sigma_- \otimes \openone)  - \frac{1}{2} \Big\{ \varrho_{SA} , (\sigma_-  \otimes \openone) (\sigma_+  \otimes \openone) \Big\} \right),
\end{aligned}
\end{equation}
where $\sigma_{+}\!=\!\sigma_-^\dagger\!=\!\ket{1}\bra{0}$ are the spin raising and lowering operators, while $\gamma$ and $\Gamma$ fix the dissipation rates. The inverse temperature of the bath is given by $\beta=\tfrac{1}{2\omega_S}\ln\left(\tfrac{\gamma}{\Gamma}\right)$. The expression $\mathcal{H}_I$ of the interaction captures a broad range of important physical processes, including energy and excitation preserving models. The dynamics of the reduced state of the system 
\begin{equation}
\varrho_S = \text{Tr}_A \left[ \varrho_{SA} \right],
\end{equation}
provides a physically legitimate evolution, crucially however, it is no longer Markovian due to the strong interaction with $A$. We remark similar settings have been explored in the literature~\cite{CampbellPRA2010, DeffnerPRL2015}, and moreover, its general validity was recently examined~\cite{ManiscalcoPRA}.

Our model therefore provides a versatile setting to explore the interrelation between the establishment of strong correlations, both quantum and classical, non-Markovian dynamics, and the behavior of the entropy production. Indeed, our model not only allows to seamlessly move from a Markovian to a non-Markovian picture, but also allows for the study of non-equilibrium steady states with respect to the bath and clearly establish a relation with the emergence of such states and the correlations shared between $S$ and $A$.

\section{Steady State Properties}
As the steady state, $\varrho_{SA}^{\infty}$, plays a crucial role in evaluating the entropy production, in this section we focus on its characteristics. By solving the LHS of Eq.~\eqref{master} set equal to zero $\varrho_{SA}^{\infty}$ can be obtained fully analytically, however given its cumbersome form we do not report it here. Regardless of the explicit expressions, some properties are immediately evident, in particular the steady state is in $X$-form and independent of $J_z$, while all other parameters enter non-trivially. The presence of off-diagonal terms implies some correlations are established between the two qubits. We can examine the quantum correlations present using the entanglement of formation (EoF)
\begin{equation}
\mathcal{E}\,{=}\,h\left(\frac{1}{2}\left[1+\sqrt{1-\mathcal{C}^2}\right]\right),
\end{equation}
where $h(x)\,{=}\,-x\text{log}_2 x - (1-x)\text{log}_2 (1-x)$ is the binary entropy function and $\mathcal{C}$ is the concurrence of the state. The latter is an equally valid entanglement measure and can be found in terms of the eigenvalues $\lambda_1\,{\geq}\, \lambda_{2,3,4}$ of the spin-flipped density matrix $\rho_{AB}(\sigma_y\otimes\sigma_y)\rho^*_{AB} (\sigma_y\otimes\sigma_y)$ as $\mathcal{C}=\text{max}\left[0,\sqrt{\lambda_1}-\sum_{i=2}^{4}\sqrt{\lambda_i}\right]$.
A more general measure of correlations in a state is given by the quantum mutual information (MI)
\begin{equation}
\label{mutinfoeq}
{\cal I}(\rho_{12})={\cal S}(\rho_1)+{\cal S}(\rho_2)-{\cal S}(\rho_{12}), 
\end{equation}

\begin{figure}[t]
{\bf (a)}\\
\includegraphics[width=0.85\columnwidth]{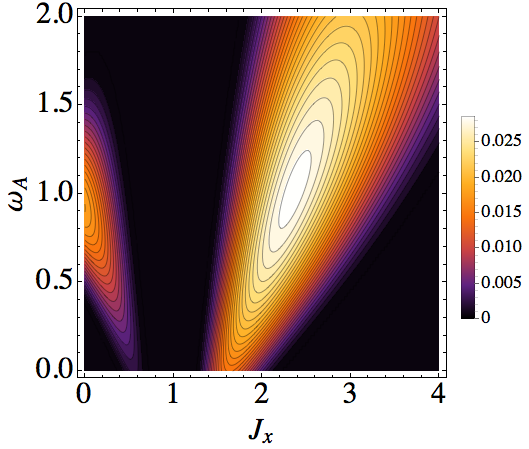}\\
{\bf (b)}\\
\includegraphics[width=0.85\columnwidth]{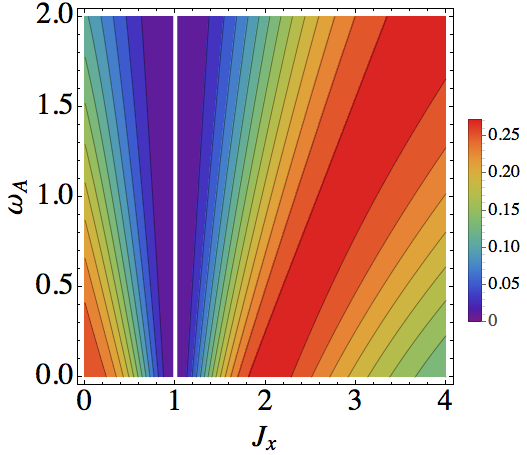}\\
{\bf (c)}\\
\includegraphics[width=0.8\columnwidth]{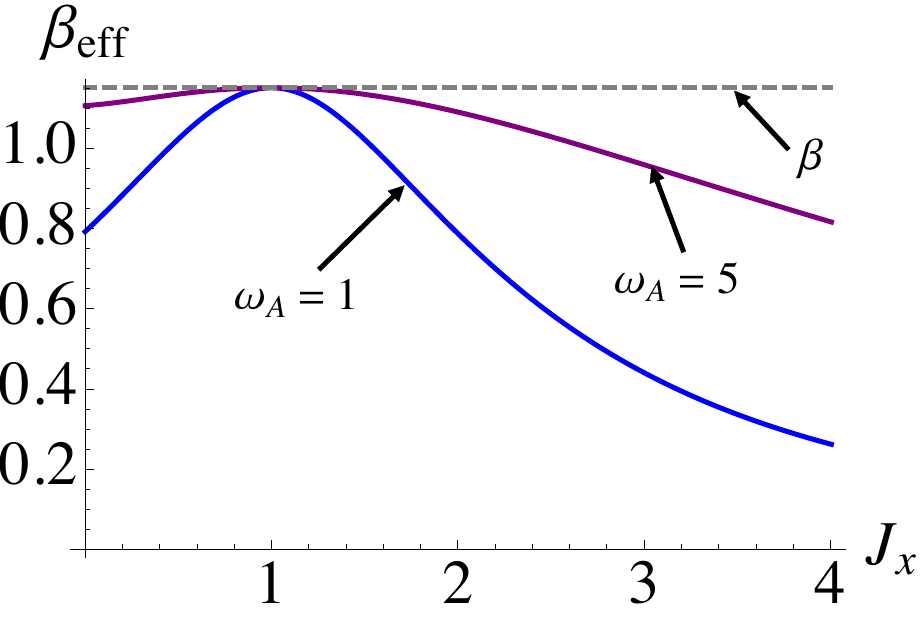}\\
\caption{{\bf (a)} Steady state EoF and {\bf (b)} MI (in base 2) versus the frequency, $\omega_A$ of the ancillary system $A$ and coupling strength $J_x$. {\bf (c)} Effective inverse temperature, $\beta_\text{eff}$ of the system $S$. In all panels we have fixed $\gamma=10$, $\Gamma = 1$, $J_y=1$, $J_z=0$, and $\omega_S=1$.}
\label{fig1}
\end{figure}

\noindent where ${\cal S}(\rho)\,{=}\,{-}\text{Tr}[\rho\log\rho]$ is the von Neumann entropy of a generic state $\rho$. This measure accounts for all correlations both classical and quantum. Fixing $J_y\!=\!1$ and $\omega_S\!=\!1$, in Fig.~\ref{fig1} {\bf (a)} and {\bf (b)} we find that both quantifiers depend non-trivially on the frequency detuning between the two spins and the coupling strength $J_x$. Immediately we see that there are significant parameter ranges where the steady state exhibits no entanglement, while the MI exhibits a markedly different dependence on the parameters, in particular, only being identically zero when $J_x\!=\!J_y$, as highlighted by the vertical white line. For other interaction strengths it is clear that the presence of correlations indicates that the strong coupling between $S$ and $A$ results in a non-equilibrium steady state with respect to the bath. As expected, the actual steady state compares to a thermal ansatz of $S\!+\!A$ given by a canonical Gibbs state, determined by the total Hamiltonian $\mathcal{H}_S + \mathcal{H}_A +\mathcal{H}_I$ and the inverse temperature of the bath $\beta$ only for $J_x\!=\!J_y$ and $\gamma\!\approx \!\Gamma$, which corresponds to an excitation preserving interaction in the limiting case of infinite temperature for the bath. As we show in the following, for other parameter choices we find that additional characteristic temperatures can emerge from $\varrho_{SA}^{\infty}$.

Due to the $X$-shape of the steady state, the reduced state of $S$, i.e. $\varrho_S^{\infty}\!=\!\text{Tr}_A[\varrho_{SA}^{\infty}]$ is diagonal for all parameter values, and therefore we can define an effective temperature, $\beta_\text{eff}$, for the system. In Fig.~\ref{fig1} {\bf (c)} we examine how the strength of the coupling between $S$ and $A$ and their relative detuning affects $\beta_\text{eff}$. When correlations are present in the steady state, as captured by the MI, the effective temperature that the system reaches is higher than that of of the bath and the discrepancy is enhanced for smaller values of $\omega_A$, corresponding to the fact that a transfer of thermal excitations to the system is favoured. Interestingly, for $J_x\!=\!J_y$ we find that $S$ thermalizes with the environment regardless of the relative detuning between the two qubits. For this value of coupling the steady state is diagonal and takes the form
\begin{equation}
\label{diagonalss}
\varrho_{SA}^{\infty} = \frac{\mathrm{e}^{-\beta \mathcal{H}_S}}{\mathcal{Z}_S} \otimes \frac{\mathrm{e}^{-\tilde{\beta} \mathcal{H}_A}}{\mathcal{Z}_A},
\end{equation}
where $\tilde{\beta}=(\beta \omega_S)/\omega_A$. The factorized form of this state puts into clear evidence the vanishing MI shown in Fig.~\ref{fig1} {\bf (b)}. It is worth stressing at this point that although $S$ is in thermal equilibrium with the bath for $J_x\!=\!J_y$, by virtue of the interaction, a second characteristic temperature emerges for $A$. The appearance of the two distinct parameters $\beta$ and $\tilde{\beta}$ can be understood as follows. For $J_x\!=\!J_y$ we have an excitation preserving interaction and the temperature associated to each system is naturally introduced by imposing a detailed balance condition for the energy exchange. For the system $S$ we have $\Gamma=\gamma \exp{(-\beta \omega_S)}$ to be compared with $\Gamma_{\mathrm{eff}}=\gamma_{\mathrm{eff}} \exp{(-\beta \omega_A)}$ for qubit $A$. Note that for this special choice of coupling the stationary state is left invariant by the Hamiltonian and the dissipative contribution separately. While the overall state can exhibit correlations, which for  $J_x\!\not =\!J_y$ can also amount to entanglement, the local state of $S$ and $A$ are in Gibbs form. Indeed the interplay between global entanglement and local thermal states appears to be a typical quantum feature \cite{Kaufman2016a}, and we remark there is evidence of a thermodynamic role played by (quantum) correlations~\cite{HuberNJP, MossyNJP, PaternostroArXiv, GooldArXiv}. Therefore, although $S$ always reaches a canonical Gibbs state at {\it some} temperature regardless of the particular parameter choice, the interaction can introduce other characteristic temperatures to the overall system. Thus, while an expression for the entropy production can be found that is meaningful, the emergence of the additional temperatures makes discussing any thermodynamic aspect more difficult. In this regard, the choice of $J_x\!=\!J_y$ and $\omega_S\!=\!\omega_A$ is special as the model retains the unique temperature defined by the bath, $\beta$. In what follows we will restrict to this setting to allow for a more meaningful and consistent thermodynamic interpretation.

\section{Entropy Production}
\subsection{Preliminaries}
\label{entropy}
Let us consider a system in contact with a bath at inverse temperature $\beta$, with which it can exchange heat. The irreversible contribution to the
entropy production for a given transformation is then defined as~{\cite{Sagawa1}}
\begin{eqnarray}
  \langle \Sigma \rangle & = & \Delta S- \beta \Delta Q,  \label{eq:irr}
\end{eqnarray}
where $\Delta S$ is the change in entropy of the system and $\Delta Q$ denotes the mean exchanged heat, so that $\langle \Sigma \rangle$ indeed provides the contribution in entropy change which cannot be traced back to a reversible heat flow. Assuming as initial and final times of the transformation zero and $t$, respectively, and defining $Q= \tmop{Tr} \rho H$, where $H$ is the system Hamiltonian, this expression can be equivalently rewritten as
\begin{eqnarray}
  \langle \Sigma \rangle & = & S ( \rho (0)|| \rho_{\beta} ) -S ( \rho (t)||
  \rho_{\beta} ) ,  \label{eq:irr1}
\end{eqnarray}
where $\rho_{\beta}$ denotes a Gibbs state for the system at inverse temperature $\beta$, and we have introduce the Umegaki's quantum relative entropy~{\cite{Nielsen2000}}
\begin{eqnarray}
  S ( \rho ||w) & = & \tmop{Tr}   \rho \log   \rho - \tmop{Tr}   \rho \log  w.
  \nonumber
\end{eqnarray}
If the dynamics of the system is given by a collection of time dependent completely positive trace preserving maps $\{\Phi (t,0)\}_{t}$, admitting $\rho_{\beta}$ as an invariant state, the irreversible entropy production as defined by Eq.~(\ref{eq:irr1}) is a positive quantity, in accordance to the second law. One can further consider the quantity
\begin{eqnarray}
  \sigma (t) & = & - \frac{\mathd}{\tmop{dt}} S ( \rho (t)|| \rho_{\beta} ) , 
  \label{eq:rate1}
\end{eqnarray}
which can be naturally interpreted as the (instantaneous) entropy production rate. Consider the case in which the collection of completely positive trace preserving maps is $P$-divisible, in the sense that the following composition law holds
\begin{eqnarray}
\Phi (t,0) & = & \Phi (t,s) \Phi (s,0),  \hspace{2em} t \geqslant s \geqslant 0, 
\label{eq:pdiv}
\end{eqnarray}
with $\Phi (t,s)$ a positive map $\forall$ $t \geqslant s \geqslant 0$, and where by definition
\begin{eqnarray}
  \Phi (t,0) \rho (0) & = & \rho (t) . \nonumber
\end{eqnarray}
Due to the fact that the relative entropy is a contraction under the action of a completely positive trace preserving map~\cite{Lindblad1975a}, and as recently shown also for a positive trace preserving map~{\cite{Reeb2017a}}, in this case also the entropy production is a positive quantity. 

As already observed in Ref.~{\cite{Spohn}} considering the special case of quantum dynamical semigroups, the very existence of an invariant state of the dynamics, say $\bar{\rho}$, not necessarily in Gibbs form, is sufficient to introduce via
\begin{eqnarray}
  \langle \bar{\Sigma} \rangle & = & S ( \rho (0)|| \bar{\rho} ) -S ( \rho
  (t)|| \bar{\rho} ) ,  \label{eq:irr2}
\end{eqnarray}
a quantifier of entropy production which is always positive, and whose associated entropy production rate $\bar{\sigma} (t)$ is also positive provided the dynamics is $P$-divisible according to Eq.~(\ref{eq:pdiv}). For the special case of a quantum dynamical semigroup with generator, $\mathcal{G}$, the entropy production rate is given by the explicit expression
\begin{eqnarray}
  \bar{\sigma} (t) & = & \tmop{Tr} \{ \mathcal{G} [ \rho (t)]( \log  
  \bar{\rho} - \log   \rho (t))\} ,  \label{eq:rate2}
\end{eqnarray}
whose positivity, following from the divisibility of the dynamics, is also known as Spohn's inequality~\cite{Spohn, Alicki1979}
\begin{eqnarray}
  \tmop{Tr} \{ \mathcal{G} [ \rho (t)]( \log   \bar{\rho} - \log   \rho (t))\}
  & \geqslant & 0. \nonumber
\end{eqnarray}
Both definitions for the entropy production rate provide convex functions of the system state, thus ensuring stability, and they are positive in the presence of a $P$-divisible dynamics. However, only $\sigma (t)$ defined in Eq.~(\ref{eq:rate1}) via its relation to Eq.~(\ref{eq:irr1}), and therefore heat transfer, can be directly connected to a thermodynamic interpretation.

Turning our attention to the dynamics, recently different notions of non-Markovianity have been introduced, related to a notion of divisibility of the quantum dynamical map or motivated by information backflow between system and environment, see~\cite{Breuer2012a, Rivas2014a, BreuerRMP, Devega2017a} for recent reviews. The definition of non-Markovianity related to divisibility was originally conceived in terms of $CP$-divisibility, that is asking that $\Phi (t,s)$ in Eq.~(\ref{eq:pdiv}) is a completely positive map {\cite{Rivas2010a}}. On the other hand, in order to connect the presence of memory effects with an information exchange between system and environment, it has been suggested to consider the behavior in time of the trace distance between time evolved distinct initial system states given by
\begin{equation}
\label{tracedist}
  D ( \rho_{1} ( t ) , \rho_{2} ( t ) ) = \frac{1}{2} \| \rho_{1} ( t ) - \rho_{2} ( t ) \| ,
\end{equation}
where the trace norm of an operator $\| A \| = \tmop{Tr} \sqrt{A^{\dag} A}$ has been introduced, reducing to the sum of the modulus of its eigenvalues for a self-adjoint operator. Revivals in time of the trace distance, Eq.~(\ref{tracedist}), for at least a pair of initial states is then assumed as a definition of non-Markovian dynamics \cite{Breuerprl}. Despite the difference in the formulation, it has been realized that there is a close connection between the notions of divisibility and information back-flow. In particular it has been shown {\cite{Chruscinski2011a,Wissmann2015a,Breuer2018a}} that the trace distance criterion is equivalent to $P$-divisibility, provided the map is invertible as a linear transformation. Thus, it is natural to study the behavior of the entropy production rate, and in particular its sign, and compare it with the assessment of non-Markovianity, without necessarily using it as a new definition of non-Markovian dynamics {\cite{Chen2017a}}.

\subsection{Dynamical behavior}
We next study the behavior of the entropy production rate in our model, which provides a simple controlled way to go from a Markovian to a non-Markovian dynamics. As stressed previously, we will restrict to the case of $J_x\!=\!J_y$ and $\omega_S=\omega_A$ where the steady state takes the form shown in Eq.~\eqref{diagonalss} with $\tilde{\beta}=\beta$. For this model an analytic treatment is feasible as shown in the Appendix, and in particular one can provide the equations describing the reduced dynamics in the form of a time-convolutionless master equation. 
\begin{figure}[t]
{\bf (a)}\\
\includegraphics[width=0.85\columnwidth]{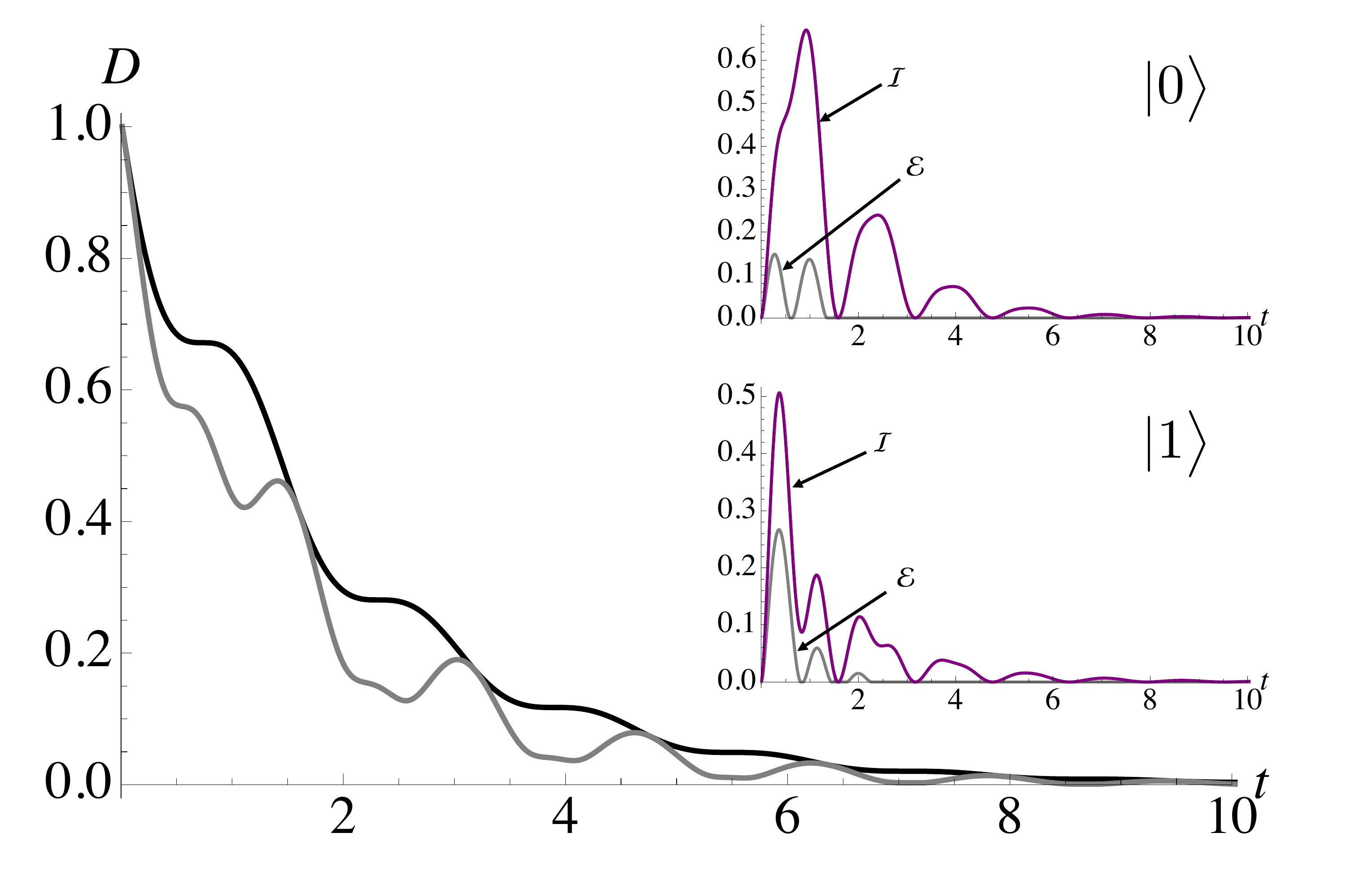}\\
{\bf (b)}\\
\includegraphics[width=0.85\columnwidth]{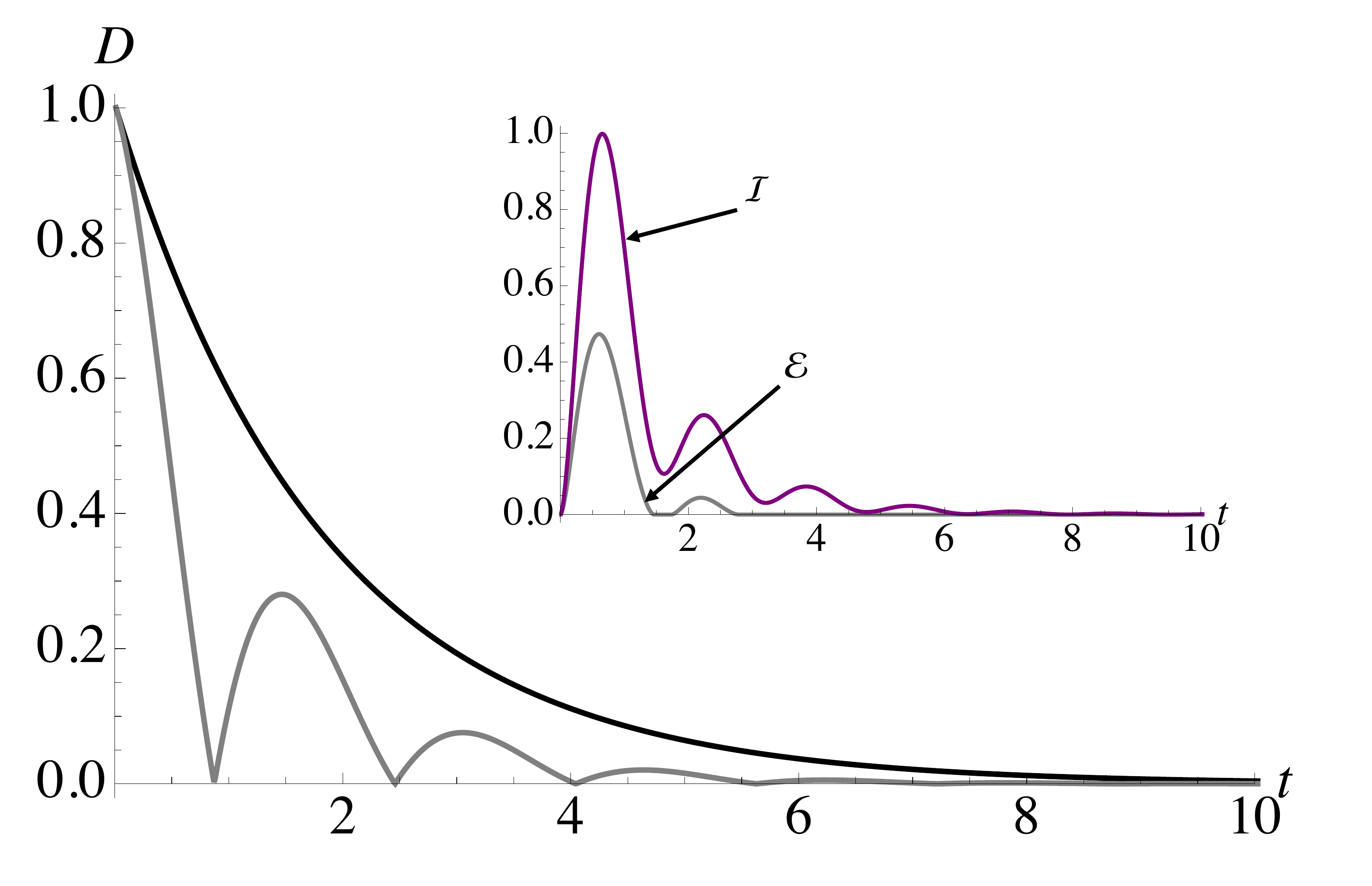}\\
\caption{Trace distance for the joint $S\!+\!A$ system (upper, black) and reduced system $S$ (lower, gray) setting $\varrho_S(0)$ to be {\bf (a)} $\{ \ket{0}\bra{0},\ket{1}\bra{1} \}$. {\bf (b)} $\{ \ket{+}\bra{+},\ket{-}\bra{-} \}$. {\it Insets:} MI (upper, purple) and EoF (lower, gray). In both panels $\varrho_A(0) = \ket{+}\bra{+}$ and we have fixed $\gamma=10 \Gamma = 1$, $J_x=J_y=1$, $J_z=0$, and $\omega_S=\omega_A=1$.}
\label{fig2}
\end{figure}

While the combined system and ancilla state obeys a semigroup dynamics described by Eq.~(\ref{master}), the reduced dynamics becomes non-Markovian once the interaction with the ancillary system is switched on. We see this clearly in Fig.~\ref{fig2} where we show the trace distance for $S\!+\!A$ (black) and the reduced system $S$ (gray) for two different pairs of orthogonal initial states. While the total state results in a monotonically decreasing behavior, since the dynamics is strictly Markovian, the interaction can lead to points of inflection/plateaus, cf. panel {\bf (a)}. In contrast, the trace distance for $S$ clearly shows the non-Markovian nature of the dynamics. In additional we see from the insets that despite the steady state exhibiting zero EoF and MI, nevertheless dynamically significant amounts of correlations can be established. As we will see, these correlations contribute non-trivially to the entropy production.

\begin{figure}[t]
\includegraphics[width=0.85\columnwidth]{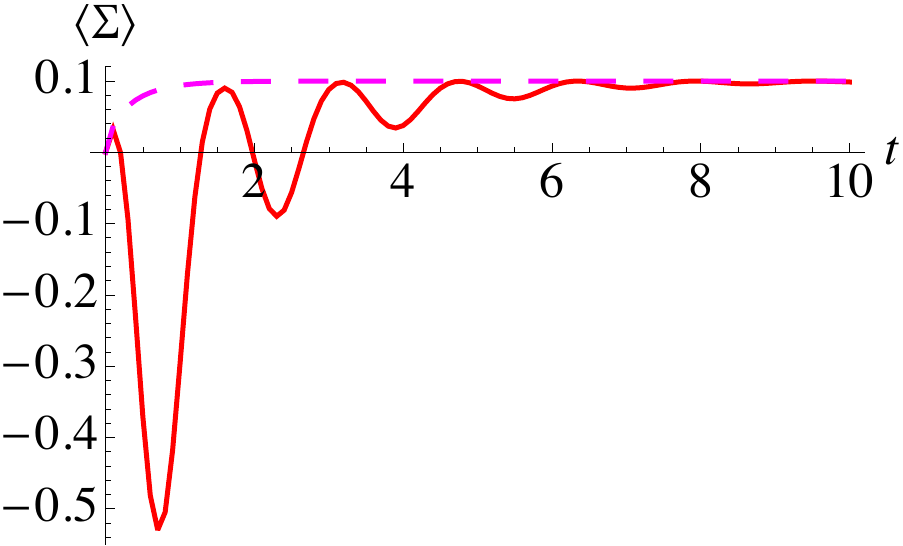}\\
\caption{Behavior of entropy production, $\langle \Sigma \rangle$, of system $S$ for non-Markovian dynamics (solid, red) compared with the corresponding Markovian case when the $S$-$A$ coupling is switched off (dashed, magenta). Here, $\varrho_S(0) = \ket{1}\bra{1}$, $\varrho_A(0) = \ket{+}\bra{+}$ and we have fixed $\gamma=10 \Gamma = 1$, $J_x=J_y=1$, $J_z=0$, and $\omega_S=\omega_A=1$. In the long time limit the two entropy productions coincide.}
\label{fig3}
\end{figure}

For $S+A$ the entropy production rate $\bar{\sigma}_{\tmop{SA}} (t)$ is known to be positive and given in particular by the expression
\begin{equation}
\bar{\sigma}_{\tmop{SA}} (t) = \tmop{Tr} \{ \mathcal{L} [ \varrho_{SA} (t)]( \log \varrho^{\infty}_{SA} - \log \varrho_{SA} (t))\} .  
\label{SAentprod}
\end{equation}
Similarly, the entropy production rate for the reduced system $S$, is captured via
\begin{equation}
\begin{aligned}
  \bar{\sigma}_{S} (t)=& - \frac{\mathd}{\tmop{dt}} S ( \varrho_{S} (t)|| \tmop{Tr}_{A}  \varrho^{\infty}_{SA} ) \\
                                 =& - \frac{\mathd}{\tmop{dt}} S ( \varrho_{S} (t)|| \rho_{\beta} ) = \sigma_{S} (t).
\label{Sentprod}
\end{aligned}
\end{equation}
Notice that Eq.~\eqref{Sentprod} retains a clear thermodynamic meaning since in this case, due to the form of coupling, the system thermalizes with the bath. In fact the long-time entropy production, $\langle \Sigma \rangle$, for $S$ is a positive quantity and, furthermore, achieves the same value regardless of whether the underlying dynamics is Markovian ($\mathcal{H}_I\!=\!0$) or non-Markovian ($\mathcal{H}_I\!\neq\!0$). However, in line with the results of Ref.~\cite{MarcantoniSciRep}, we see the entropy production can become transiently negative in the non-Markovian case, as shown in Fig.~\ref{fig3}. 

In the present setting, thanks to the general identity
\begin{eqnarray}
\label{sss}
  S ( \rho_{SA} | | \tau_{S} \otimes w_{A} ) & = & \mathcal{I} ( \rho_{SA} )
  +S ( \rho_{S} | | \tau_{S} ) +S ( \rho_{A} | | w_{A} ), 
\end{eqnarray}
relying on the structure of the equilibrium state Eq.~\eqref{diagonalss}, we find a simple relation between the various contributions to the entropy production rate for $S+A$ and the establishment of correlations
\begin{equation}
\label{eq:sigmasum}
\bar{\sigma}_{\tmop{SA}} (t) = \bar{\sigma}_{\tmop{S}} (t) +\bar{\sigma}_{\tmop{A}} (t) -  \frac{\mathd}{\tmop{dt}} {\cal I}(\rho_{SA}(t)).
\end{equation} 
Such a relation immediately puts into evidence the non-trivial role that the dynamical build-up of correlations between the constituents of the total system plays in the proper thermodynamical characterization of the process, both in open and closed system settings~\cite{EspositoNJP, CampisiJPA}. 

\begin{figure}[t]
{\bf (a)}\\
\includegraphics[width=0.85\columnwidth]{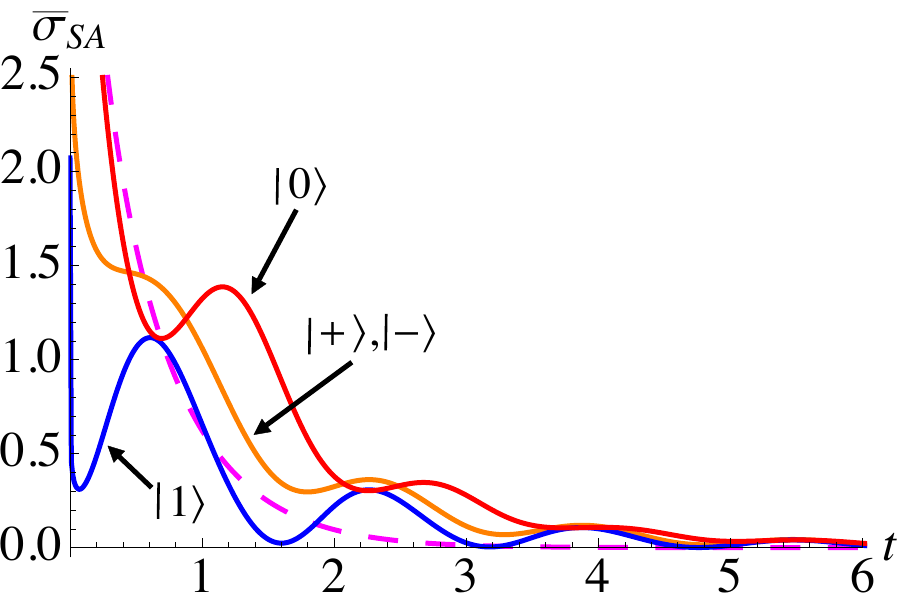}\\
{\bf (b)}\\
\includegraphics[width=0.85\columnwidth]{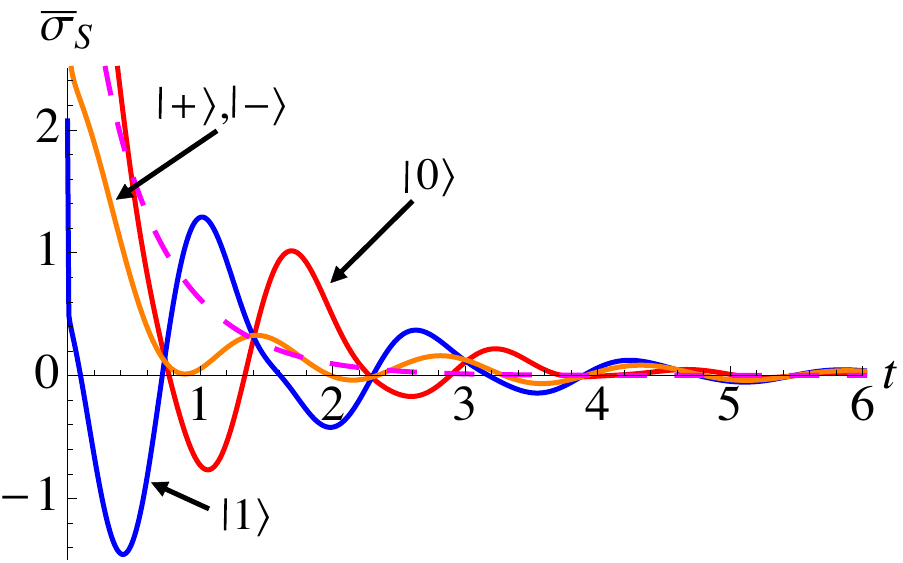}\\
\caption{{\bf (a)} Entropy production rate for $S+A$ Eq.~\eqref{SAentprod}. {\bf (b)} Entropy production rate for the reduced system $S$ Eq.~\eqref{Sentprod}. In both panels $\varrho_A(0)\!=\!\ket{+}\bra{+}$ and we have fixed $\gamma\!=\!10 \Gamma\!=\!1$, $J_x\!=\!J_y\!=\!1$, $J_z\!=\!0$, and $\omega_S\!=\!\omega_A\!=\!1$. In both panels the dashed magenta curve corresponds to the respective entropy production rate when $\mathcal{H}_I=0$ and $\varrho_S(0)=\ket{0}\bra{0}$. The curve labels correspond to the initial states of $S$.}
\label{fig4}
\end{figure}

For zero interaction between the qubits, the entropy production rate, Eq.~\eqref{SAentprod}, is a strictly monotonically decreasing function as shown by the dashed curves in Fig.~\ref{fig4}. We examine the effect that the interaction term has on $\bar{\sigma}_{\tmop{SA}}$ in Fig.~\ref{fig4} {\bf (a)} for various initial states of $S$ when $A$ is initialized in $\ket{+}$.  While in line with Markovian dynamics $\bar{\sigma}_{SA}\geq 0$ for all parameter choices, the general behavior deviates significantly from the Markovian case and we see strong oscillations occurring. In Fig.~\ref{fig4} {\bf (b)} we examine the entropy production rate for the system, $S$, for the same parameters. We now see that the $\bar{\sigma}_{S}$ can dynamically become negative and the periods during which this happens are closely related to when oscillations occur in $\bar{\sigma}_{SA}$. Such a behavior is consistent with other studies assessing entropy production rates in non-Markovian settings~\cite{MarcantoniSciRep, PatiPRA}, where the stationary state is always in Gibbs form. 

Connecting this to non-Markovianity as described by an information back-flow between system and environment, as discussed in Sec.~\ref{entropy}, this exchange in information can be traced back to a change in time of the distinguishability between distinct system states. In the present setting it is natural to consider the relative entropy as a quantifier of the distinguishability between system states. We can therefore study $S(\varrho^1_S(t)~\|~\varrho^2_S(t))$ for different choices of $\varrho^{1,2}_S(0)$. For the special case of $\varrho^{2}_S(0)=\varrho_S^{\infty}$ one describes the entropy production rate as a change of distinguishability, and therefore an indicator of non-Markovianity, thus, under this viewpoint, negative entropy production rates and non-Markovianity are directly related. However, considering the behavior of the trace distance shown in Fig.~\ref{fig2}, the revivals, which indicate periods of non-Markovianity, are not in one-to-one correspondence with the negative entropy production rates for $S$ in Fig.~\ref{fig4} {\bf (b)}. Therefore, periods of non-Markovian dynamics under one figure of merit do not directly imply a negative entropy production rate.

We can therefore conclude that a violation of $P$-divisibility alone can lead to dynamically negative entropy production rates. However this must be caveated, for $J_x\!=\!J_y$ there are initial states of systems $A$ and $S$ for which $\sigma_S\geq0$ at all times. For example, setting $\varrho_A(0)\!=\!\varrho_S(0)\!=\!\tfrac{1}{2}\openone$ leads to a positive entropy production rate for $S$ at all times despite its dynamics being non-Markovian. Thus, we have that satisfying the $P$-divisibility property is a sufficient but not necessary condition for a positive entropy production rate, since there are non $P$-divisible dynamical maps that still lead to $\sigma_S\geq0$ $\forall t$ for particular initial conditions.

\section{Conclusions}
We have examined the steady state correlation properties and dynamical entropy production in a versatile setting, consisting of a single qubit, $S$, embedded in a Markovian bath interacting with a clean ancillary qubit, $A$. While the overall dynamics remains Markovian, this is no longer true for the reduced dynamics of the system due to its interaction with $A$. We have shown that the interaction can lead to strongly correlated steady states for the joint system. Examining the resulting non-equilibrium steady state we found that when such correlations are present, $S$ exhibits an effective temperature which is higher than the bath. We highlight the special choice of an excitation preserving interaction, which ensured no correlations were present in the steady state. In this setting $S$ reached thermal equilibrium with the bath and therefore allowed for a meaningful assessment of the thermodynamic features of the dynamics. By computing the entropy production rate we showed that the non-Markovianity induced by the interaction with $A$ could lead to negative entropy production rates, while the overall entropy production was still strictly positive.

Our study reveals the highly non-trivial role that the establishment of correlations plays in the thermodynamic characterization of quantum systems. At the level of the system, we have shown that the absence of quantum correlations, i.e. entanglement, is not sufficient to ensure meaningful thermodynamic quantities. Rather, {\it any} correlations can greatly complicate both the dynamical and steady state properties. In line with previous studies we have shown that witnessing a negative entropy production rate for the system due to a non-Markovian dynamic does not imply a violation of the second law. For instance, Ref.~\cite{MarcantoniSciRep} insisted that one should keep track of the entropy changes from both the system and the environment, our study goes further by showing that one must also take into account the correlations established between the two, i.e. Eq.~\eqref{eq:sigmasum}. The versatile nature of our model further reveals that, while clearly interrelated, the establishment of correlations, non-Markovian dynamics, and (negative) entropy productions rates is quite complex. In particular, as our study has revealed, non-Markovianity appears to be sufficient, but not necessary, to realize negative entropy production rates.

\acknowledgements
We are grateful to Matteo Brunelli and Sebastian Deffner for enlightening discussions and comments. BV acknowledges support from the EU Collaborative project QuProCS (grant agreement 641277) and FFABR. 

\section*{Appendix}
In the following appendix we provide the analytic solution of the model Eq.~\eqref{master} in the particular case when $J_{x}=J_{y}=\frac{J}{8}$,  $J_{z}=0$, and $\omega_{S}=\omega_{A}=\omega$. The set of differential equations for the elements of density matrix $\varrho_{SA}$ are given by
\begin{eqnarray*}
\dot{\varrho}_{SA}^{0000}\left(t\right)&=&-\gamma\varrho_{SA}^{0000}\left(t\right)+\Gamma\varrho_{SA}^{1010}\left(t\right)\\
\dot{\varrho}_{SA}^{0001}\left(t\right)&=&i\,\frac{J}{4}\,\varrho_{SA}^{0010}\left(t\right)-\left(2i\,\omega+\gamma\right)\varrho_{SA}^{0001}\left(t\right)+\Gamma\varrho_{SA}^{1011}\left(t\right)\\
\dot{\varrho}_{SA}^{0010}\left(t\right)&=&-\left(\frac{1}{2}(\gamma+\Gamma)+2i\omega\right)\varrho_{SA}^{0010}\left(t\right)+i\,\frac{J}{4}\varrho_{SA}^{0001}\left(t\right)\\
\dot{\varrho}_{SA}^{0011}\left(t\right)&=&-\frac{1}{2}\left(\gamma+\Gamma+8i\omega\right)\varrho_{SA}^{0011}\left(t\right)\\
\dot{\varrho}_{SA}^{0101}\left(t\right)&=&-\left(\gamma+\Gamma\right)\varrho_{SA}^{0101}\left(t\right)+\Gamma\left(1-\varrho_{SA}^{0000}\left(t\right)-\varrho_{SA}^{1010}\left(t\right)\right)\\ 
                                              &&-i\,\frac{J}{4}\left(\varrho_{SA}^{0110*}\left(t\right)-\varrho_{SA}^{0110}\left(t\right)\right)\\
\dot{\varrho}_{SA}^{0110}\left(t\right)&=&i\,\frac{J}{4}\left(\varrho_{SA}^{0101}\left(t\right)-\varrho_{SA}^{1010}\left(t\right)\right)-\frac{1}{2}\left(\gamma+\Gamma\right)\varrho_{SA}^{0110}\left(t\right)\\
\dot{\varrho}_{SA}^{0111}\left(t\right)&=&-\left(\frac{1}{2}\left(\gamma+\Gamma\right)+2i\omega\right)\varrho_{SA}^{0111}\left(t\right)-i\,\frac{J}{4}\varrho_{SA}^{1011}\left(t\right)\\
\end{eqnarray*}
\begin{eqnarray*}
\dot{\varrho}_{SA}^{1010}\left(t\right)&=&\gamma\varrho_{SA}^{0000}\left(t\right)-\Gamma\varrho_{SA}^{1010}\left(t\right)+i\,\frac{J}{4}\left(\varrho_{SA}^{0110*}\left(t\right)-\varrho_{SA}^{0110}\left(t\right)\right)\\
\dot{\varrho}_{SA}^{1011}\left(t\right)&=&-i\,\frac{J}{4}\,\varrho_{SA}^{0111}\left(t\right)-\left(2i\,\omega+\Gamma\right)\varrho_{SA}^{1011}\left(t\right)+\gamma\varrho_{SA}^{0001}\left(t\right)
\end{eqnarray*}

For the reduced system's dynamics the density matrix, $\varrho_{S}$, obeys the differential equations
\begin{eqnarray*}
\dot{\varrho}_{S}^{00}\left(t\right)&=&\Gamma-\left(\gamma+\Gamma\right)\varrho_{S}^{00}\left(t\right)-i\,\frac{J}{4}\left(\varrho_{SA}^{0110*}\left(t\right)-\varrho_{SA}^{0110}\left(t\right)\right)\\
\dot{\varrho}_{S}^{01}\left(t\right)&=&-\frac{1}{2}\left(\gamma+\Gamma+4i\omega\right)\varrho_{S}^{01}\left(t\right)+i\,\frac{J}{4}\left(\varrho_{SA}^{0001}\left(t\right)-\varrho_{SA}^{1011}\left(t\right)\right).
\end{eqnarray*}
For both the total $S+A$ and reduced system, the remaining density matrix elements can be readily obtained by exploiting normalization and Hermicity.

From the above equations for the reduced system we can determine the explicit form of the generator, $\mathcal{K}_{S}$, of the non-Markovian dynamics. Fixing the initial state of the ancilla to be $\varrho_{A}\left(0\right)=\frac{1}{2}\openone$ in the basis $\left\{\frac{\openone}{\sqrt{2}},\,\sigma_{-},\sigma_{+},\frac{\sigma_{z}}{\sqrt{2}}\right\}$ the generator has the form
\begin{equation*}
\begin{aligned}
\mathcal{K_{S}}\left(\varrho_{S}\right)	&=-i\left[\mathcal{H}_{S},\varrho_{S}\right]+\frac{1}{D}\left[  \gamma_{1}^{S}\left(t\right) \left(  \sigma_{z} \varrho_{S} \sigma_{z}-\varrho_{S}  \right)  \right. \\
&+ \gamma_{2}^{S}\left(t\right) \left( \sigma_{-}\varrho_{S}\sigma_{+}-\frac{1}{2}\left\{ \sigma_{+}\sigma_{-},\varrho_{S}\right\}  \right)\\
&+ \gamma_{3}^{S}\left(t\right)\left(\sigma_{+}\varrho_{S}\sigma_{-}-\frac{1}{2}\left\{ \sigma_{-}\sigma_{+},\varrho_{S}\right\} \right)\\
&+ e^{-i(2\omega t + \frac{\pi}{4})}\gamma_{4}^{S}\left(t\right)\left(\sigma_{z}\varrho_{S}\sigma_{+}-\frac{1}{2}\left\{ \sigma_{+}\sigma_{z},\varrho_{S}\right\} \right)\\
&+ e^{-i(2\omega t + \frac{\pi}{4})}\gamma_{4}^{S}\left(t\right) \left(\sigma_{+}\varrho_{S}\sigma_{z}-\frac{1}{2}\left\{ \sigma_{z}\sigma_{+},\varrho_{S}\right\} \right)\\
&+ e^{i(2\omega t + \frac{\pi}{4})}\gamma_{4}^{S}\left(t\right)\left(\sigma_{z}\varrho_{S}\sigma_{-}-\frac{1}{2}\left\{ \sigma_{-}\sigma_{z},\varrho_{S}\right\} \right)\\
&+\left. e^{i(2\omega t + \frac{\pi}{4})}\gamma_{4}^{S}\left(t\right)\left(\sigma_{-}\varrho_{S}\sigma_{z}-\frac{1}{2}\left\{ \sigma_{z}\sigma_{-},\varrho_{S}\right\} \right)\right]
\end{aligned}
\end{equation*}
where
\begin{eqnarray*}
\gamma_{1}^{S}\left(t\right)&=&4\dot{\lambda_{1}}\left(\lambda_{3}-\lambda_{4}\right)-2\lambda_{1}\left( \dot{\lambda_{3}}  -\dot{\lambda_{4}}\right)\\
\gamma_{2}^{S}\left(t\right)&=&-2\lambda_{1}\left(\lambda_{4}\left(\dot{\lambda_{2}}+\dot{\lambda_{3}}\right)+\left( \lambda_{2}-1 \right)\left(\dot{\lambda_{3}}-\dot{\lambda_{4}}\right)-\lambda_{3}\left( \dot{\lambda_{2}}+\dot{\lambda_{4}}\right)\right)\\
\gamma_{3}^{S}\left(t\right)&=&4\lambda_{1}\left(\dot{\lambda_{3}}-\dot{\lambda_{4}}\right)-\gamma_{2}^{S}\left(t\right)\\
\gamma_{4}^{S}\left(t\right)&=&\sqrt{2} \left(\lambda_{5}-\lambda_{2}\right)\left(\dot{\lambda_{3}}-\dot{\lambda_{4}}\right)- \sqrt{2} \lambda_{4}\left(\dot{\lambda_{2}}+\dot{\lambda_{3}}-\dot{\lambda_{5}}\right)\\
&+& \sqrt{2} \lambda_{3}\left(\dot{\lambda_{2}}+\dot{\lambda_{4}}-\dot{\lambda_{5}}\right)
\end{eqnarray*}
and $D=4\lambda_{1}\left(\lambda_{4}-\lambda_{3}\right)$, with $\left\{ \lambda_{i}\right\}$ being dimensionless functions defined below and $\left\{ \dot{\lambda_{i}}\right\}$ being their time derivatives. It is clear from the generator $\mathcal{K}_{S}$, which is written here in Lindblad form, that the non-Markovian nature of the reduced dynamics arises from the time dependency of the rates $\left\{ \gamma_{j}\left(t\right)\right\}$. This time dependency is exponential, as can be seen from the explicit form of the $\left\{ \lambda_{i}\right\}$ functions that appear in $\left\{ \gamma_{j}\left(t\right)\right\}$
\begin{eqnarray*}
\lambda_{1}&=&\frac{e^{-\frac{1}{2}t\eta}\left(\sqrt{\Omega_{-}}\sinh\left(\frac{t\sqrt{\Omega_{+}}}{2\sqrt{2}}\right)-\sqrt{\Omega_{+}}\sinh\left(\frac{t\sqrt{\Omega_{-}}}{2\sqrt{2}}\right)\right)}{\sqrt{2 \Delta}}\\ \nonumber
&+& \frac{e^{-\frac{1}{2}t\eta}\left(\Sigma_{+}\cosh\left(\frac{t\sqrt{\Omega_{-}}}{2\sqrt{2}}\right)-\Sigma_{-}\cosh\left(\frac{t\sqrt{\Omega_{+}}}{2\sqrt{2}}\right)\right)}{2\sqrt{\Delta}}\\ \nonumber
\end{eqnarray*}
\begin{eqnarray*}
\lambda_{2}&=&\left(\sqrt{\Delta}\eta(\Gamma-\gamma)+\gamma \eta e^{-\frac{1}{2}t\left(\eta+\sqrt{\Delta}\right)}\left(\Sigma_{+}-\Sigma_{-}e^{\sqrt{\Delta}t}\right)\right)\Delta^{-3/2}\\ \nonumber
\end{eqnarray*}
\begin{eqnarray*}
\lambda_{3}&=&\frac{J^{2}\left(\sqrt{\Delta}(\gamma-\Gamma)+2\gamma e^{-\frac{1}{2}t\eta}\left(\eta\sinh\left(\frac{\sqrt{\Delta}t}{2}\right)-\sqrt{\Delta}\right)\right)}{\Delta^{3/2}\eta}\\ \nonumber
\end{eqnarray*}
\begin{eqnarray*}
\lambda_{4}&=&\frac{J^{2}}{\eta}\left(\frac{1}{2\Delta}(\gamma-\Gamma)\left(e^{\frac{\sqrt{\Delta}t}{2}}-1\right)^{2}e^{-\frac{1}{2}t\left(\eta+\sqrt{\Delta}\right)}-\frac{\,\lambda_{2}}{\eta}\right)\\
\end{eqnarray*}
\begin{eqnarray*}
\lambda_{5}&=&\frac{e^{-\frac{1}{2}t\eta}\left(2J^{2}\Gamma+\Delta(\Gamma-\gamma)e^{\frac{1}{2}t\eta}\right)}{\Delta\eta}\\
&-&\frac{e^{-\frac{1}{2}t\left(\eta+\sqrt{\Delta}\right)}\Gamma\left(\sqrt{\Delta}\left(1-e^{\sqrt{\Delta}t}\right)+\eta\left(1+e^{\sqrt{\Delta}t}\right)\right)}{\Delta}
\end{eqnarray*}
where $\eta =\gamma+\Gamma$, $\Delta =\eta^{2}-J^{2}$, $\Omega_{\pm} =\eta^{2}-\frac{J^{2}}{2} \pm\eta\sqrt{\Delta}$, and $\Sigma_{\pm} =\left(\eta \pm\sqrt{\Delta}\right)$.
Note that if condition $\eta>J$  is satisfied, functions $\left\{ \lambda_{i}\right\}$ are real functions $\forall i$. Consequently, functions $\left\{ \gamma_{j}\left(t\right)\right\}$ are also real functions. 

\bibliography{entropy_production}
\end{document}